\documentclass[baaa]{baaa}

\usepackage[pdftex]{hyperref}
\usepackage{subfigure}
\usepackage{natbib}
\usepackage{helvet,soul}
\usepackage[font=small]{caption}

\begin{document}


\journalvol{58}
\journalyear{2016}
\journaleditors{P. Benaglia, D.D. Carpintero, R. Gamen \& M. Lares}


\contriblanguage{1}


\contribtype{2}

\thematicarea{10}

\title{Archaeoastronomy and the orientation of old churches}


\titlerunning{Archaeoastronomy and the orientation of old churches}


\author{A. Gangui\inst{1,2}}
\authorrunning{Gangui}


\contact{gangui@iafe.uba.ar ; http://cms.iafe.uba.ar/gangui/}

\institute{Instituto de Astronomía y Física del Espacio (CONICET-UBA) \and
           Consejo Nacional de Investigaciones Científicas y Técnicas }


\abstract{ 
Cultural astronomy is an interdisciplinary area of research that studies how perceptions and concepts related to the sky are part of the worldview of a 
culture. One of its branches, archaeoastronomy, focuses on the material remains of past peoples and tries to investigate their practices and astronomical knowledge. In this
context, the orientation of Christian churches is now considered a distinctive feature of their architecture that repeats patterns from early Christian times. There is a
general tendency to align their altars in the solar range, with preference for orientations towards the east. Here we present recent data from our measurements of
astronomical orientations of old churches located in two --geographically and culturally-- very distant regions, and we discuss the results in the light of the historical and
cultural knowledge surrounding these temples.
}

\resumen{
La astronomía cultural es un área de investigación interdisciplinaria que estudia la forma en que las percepciones y conceptos sobre el cielo forman parte integrante de la
cosmovisión de una cultura. Una de sus ramas, la arqueoastronomía, se concentra en los restos materiales de los pueblos del pasado y trata de indagar sobre sus prácticas y
conocimientos astronómicos. En este marco, la orientación de las iglesias cristianas es considerada hoy un elemento distintivo de su arquitectura que repite patrones desde
época paleocristiana. Existe una tendencia general a orientar sus altares en el rango solar, con una predilección por las orientaciones cercanas al este geográfico
(equinoccio astronómico). En este trabajo presentamos datos recientes de nuestras mediciones de orientaciones astronómicas de iglesias antiguas ubicadas en dos regiones
--cultural y geográficamente-- muy distantes, y discutimos los resultados a la luz del conocimiento histórico y cultural que rodea a esos templos.
}



\maketitle


\noindent 
The study of the orientation of old churches is, along with the pyramids of Egypt, European megalithic monuments and Mesoamerican ceremonial buildings, 
among the oldest and most researched topics of
archaeoastronomy. It is known that the spatial orientation of the ancient Christian churches is one of the most salient features of their architecture. In Europe and in many
remote sites where the missionaries arrived, there is a marked tendency to orient the altars of the temples in the solar range. That is, the axis of the temple, from the
front door towards the altar, is aligned to those points on the horizon where the Sun rises on different days of the year. Among these days, there is a marked preference for
those corresponding to astronomical equinoxes, when the axes point towards geographical east \citep{mccluskey-1998}. However, even within the solar range, alignments in the
opposite direction --with the altar to the west-- are not uncommon, although they are exceptional because they do not follow the canonical pattern (see
\citep{gangui_etal_2014} and references therein).

In this paper we report on our recent measurements of astronomical orientations of old churches located in two, in principle, very different regions, both geographically and
culturally, but with the common feature of having been evangelized by missionaries of the same religious faith. The first group of temples corresponds to the churches and
chapels of Lanzarote, in the Canary Islands \citep{gangui_etal_2015}, conquered and colonized by the European population in the early fifteenth century. The second set of old
constructions is that of the Andean churches of the Arica and Parinacota region, in northern Chile \citep{gangui_guillen_pereira_2016}, an extended and difficult to travel
area that received little attention from parish priests, and where one might expect some dialogue to have taken place between the Western tradition and the local Aymara
culture in regard to the design and construction of temples within the Indian reductions. After presenting some background and a summary of the data gathered in our field
work, we will briefly discuss the results in the light of the historical and cultural knowledge surrounding these temples.

\section {Churches and chapels of Lanzarote}

Religious architecture on the island of Lanzarote began with the building of single-room modest chapels. To some of them, over time, it was added small shrines or altars in
their headers, vestries on their sides and other elements of practical use such as low barbicans bordering the atrium and calvaries. In general, these constructions were not
subjected to strict execution plans, and thus their structure was erected according to the needs of the moment. 

Of the 32 chapels and churches we measured, 12 are oriented in the northern quadrant (i.e., between azimuths 315$^{\circ}$ and 45$^{\circ}$), 2 in the western quadrant, 17
are oriented in the eastern quadrant (with 13 of them in the solar range) and only 1 in the southern quadrant (we present full data in
\citep{gangui_etal_2014,gangui_etal_2015}). Our sample is representative of the island of Lanzarote (although it is not of all the Canary Archipelago), and in it two distinct
orientations are distinguished: (i) to the north (with entrance on the leeward side, avoiding perhaps the dominant winds of the place), and (ii) eastward, with the altar of
the chapel pointing towards the eastern quadrant. The 13 monuments facing \textit{ad orientem} fall within the logic observed in other studies of orientations of churches,
but what is remarkable here is the large number oriented to the northern quadrant, falling outside of the solar range. It seems to be a case singular of Lanzarote where
practical issues (to shield from the N-NE trade winds) appear to be (strongly) combined with cultic and canonical traditions (i.e., the orientations within the solar
range).

\begin{figure}[!ht]
  \centering
  \includegraphics[width=0.45\textwidth]{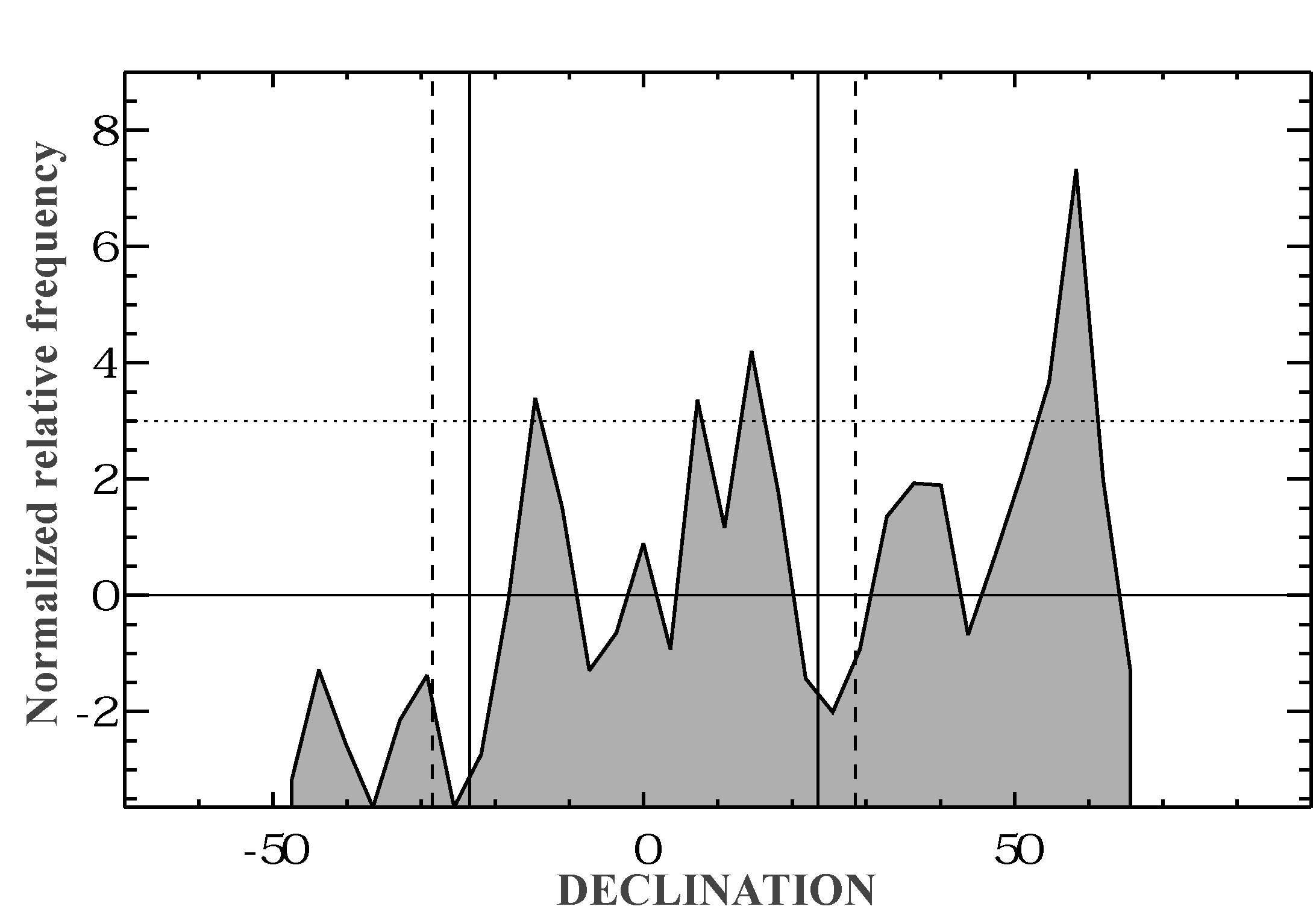}
  \caption{Declination histogram for the chapels and churches of Lanzarote. Only a few statistically significant peaks are found above the 3$\sigma$ level. Three
    statistically significant minor peaks are found in the solar range (canonical orientation). However, the highest peak, located around 58$^{\circ}$ is exceptional and
    could be associated with a accumulation peak due to orientations near the meridian line.}
  \label{decli-lanza}
\end{figure}

To better understand the above orientations, in Fig.\ref{decli-lanza} we present the declination histogram, which is independent of the geographical location and the local
topography. In the figure, continuous vertical lines represent declinations corresponding to the extreme positions of the Sun at the solstices, while the dashed vertical
lines represent the same for the Moon in major lunistices. The plot shows the astronomical declination versus the normalized relative frequency, which enables a clear and
more accurate determination of the structure of peaks. Again, the peak associated with the orientations to the north-northeast, absolutely outstanding, dominates the chart.

The particular features of the churches we measured in this island have little correlation with other studies already mentioned above. In the Iberian Peninsula and overall in
the Mediterranean, the orientation ranges are predominantly solar. In particular, the large proportion of churches oriented roughly northward we found is brand new. It is
noteworthy that a significant proportion of churches oriented in this way belong to the northwestern and central sectors of the island.

The difference between our present results for Lanzarote and those of other studies that have been completed elsewhere, leads us to look for alternatives in trying to
understand the pattern of orientations. If these monuments, in general, are not oriented according to the Sun, could it be due to such prosaic reasons as the need to orient
the porch of the constructions contrary to the dominant winds coming from the N-NE direction onto the island or otherwise protect it? Or else, could it be due to the topography
(perhaps changing over time) of different regions of the island? In any case, it seems clear that the environmental issue might be relevant.

Regarding the winds, the areas where more churches facing north-northeast have been built (with their entrance oriented towards the southern quadrant) is on the verge of El
Jable (north and center of the island), where it becomes imperative to avoid the sand driven by the wind, sometimes in raging storms as that of 1824 that buried several
villages, and that even today, despite the changes in the landscape, shows its lasting effects. Interestingly, the highest number of canonical orientations (i.e., eastward)
is found in buildings located in the northeast of the island, in the lee of the wind, in areas protected of the sand by the cliffs of Famara.

\section {Andean churches of Arica and Parinacota}

After a certain time of the Spanish settlement in the Viceroyalty of Peru, Francisco de Toledo decided to reorganize the territory and population. He paid special attention
to the trade routes of the southern Andean area, whose main objective was to organize the transport of silver from Potosi to the Pacific and also the transport of quicksilver
from Huancavelica to the high Andean mines. On the way to Arica, official maritime port for the merchandise since 1574, small villages and \textit{tambos} (buildings found
along Incan roads) with stable populations were formed. Andean churches in this region emerged in strategic locations along the route which \textit{trajinantes} marched to
transport the precious metals from Potosi to Arica beaches, particularly around the Lluta and Azapa valleys.

Religious architecture that began to materialize in the southern Andean region was characterized by elongated simple structures, rounded apses (\textit{ochavados}), stone walls
without edging, pitched and thatched roofs. In areas outside the temples there appeared catechetical crosses, arcaded atria, bell towers and small miserere chapels. In the
surrounding environments to the churches, the indigenous demands embodied in the sacred use of open spaces was clear, and formalized in a sort of cohabitation between
indigenous and European morphological ideologies.

Of the 38 measured Andean churches and chapels, 6 are oriented in the northern quadrant. There are 8 oriented in the eastern quadrant (6 of them in the solar range) and 16
oriented in the western quadrant (11 of them in the solar range, with azimuths between 245.0$^{\circ}$ and 294.6$^{\circ}$). Finally, there are 8 churches in the southern
quadrant (azimuths between 135$^{\circ}$ and 225$^{\circ}$). We think that our sample is representative of most of the temples in the region. In our data, a prominent
orientation towards the western quadrant is apparent.

We should bear in mind, however, that in many of the sites we explored geographical conditions are unique and certainly have played a role in choosing the location of the
churches; for example, those placed along rivers in some of the deep valleys (\textit{quebradas}) of the region. Moreover, apart from these riverbeds and streams, numerous
volcanoes and snowy peaks in the region can serve as points of reference --like tutelary hills or \textit{Apus} \citep{reinhard-1983}, even related to ancestor worship-- when
deciding the location and orientation of the temples, so we also checked our data to verify that possible orographic orientation.

\begin{figure}[!ht]
  \centering
  \includegraphics[width=0.45\textwidth]{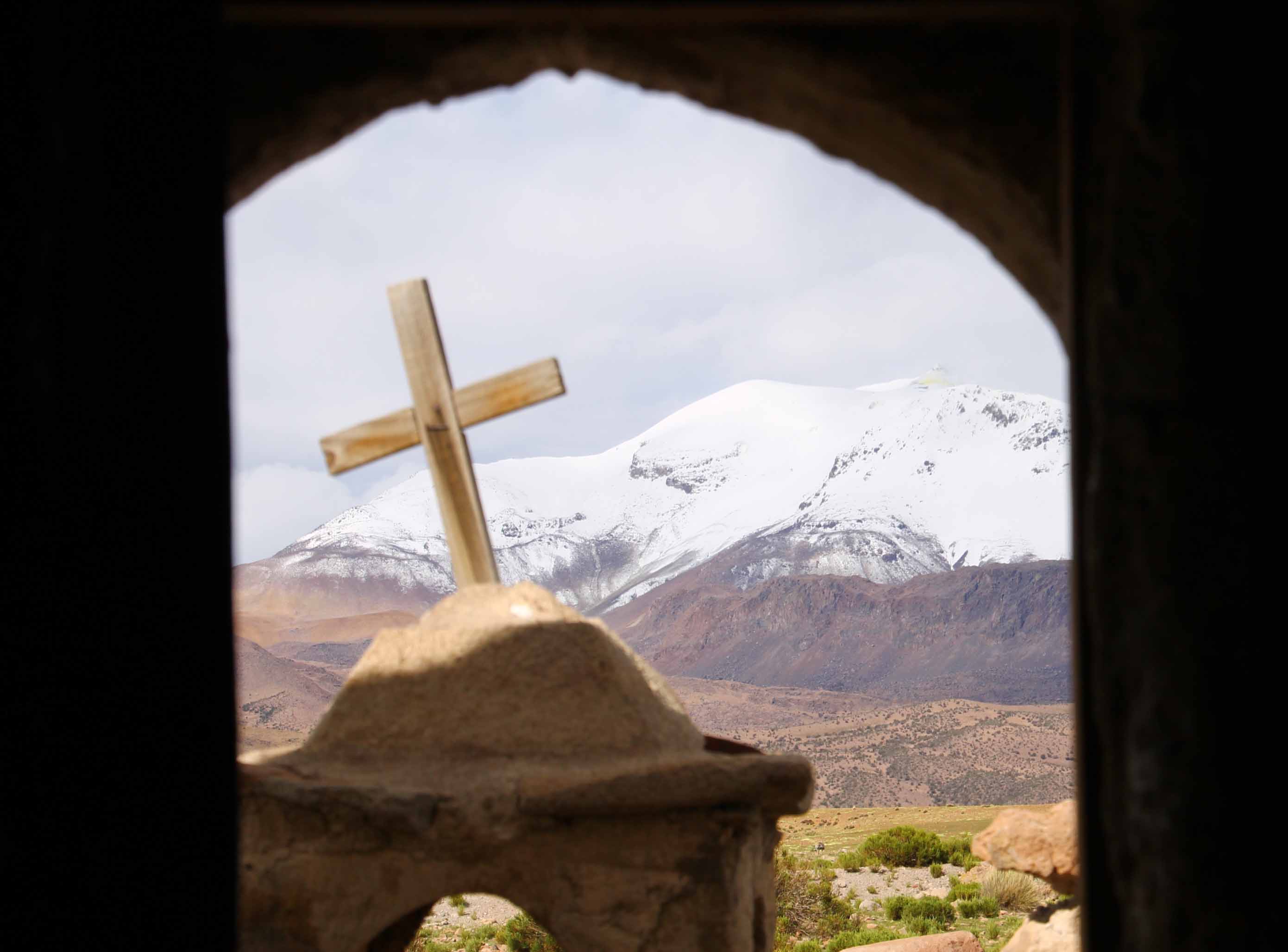}
  \caption{As one exits the chapel Virgen de la Inmaculada Concepción of Ancuta, one faces --with a high degree of precision-- the Guallatiri volcano. Photograph by A.G.}
  \label{cross-ancuta}
\end{figure}

In the case of the chapel Virgen del Carmen, in Ungallire, as one emerges from the inside, on the very front, one finds the stunning view of the volcanoes Pomerape and
Parinacota (the \textit{payachatas}); these are located with azimuths just a few degrees on either side of the main axis of the construction. We verified something similar
with the churches Virgen de la Inmaculada Concepción, of the two villages Guallatire and Ancuta: according to our measurements, the Guallatiri volcano is nearly aligned with
the axes of these churches but, in both cases, as it happened with the Ungallire chapel, the location of the volcano is in front of the buildings (Fig.\ref{cross-ancuta}).

Based on our analysis we think that often, although not always, of course (more examples in our recently submitted \citep{gangui_guillen_pereira_2015}), the characteristics
of topography and surrounding landscape in each temple site prevailed over European and colonial traditions regarding the orientation of the churches' main axes, a fact that
--in some cases-- brings to mind more the Aymara worship \citep{bouysse-cassagne-1987} than the XVIth century \textit{Instrucciones de la fábrica y del ajuar eclesiásticos}
written by Cardinal Carlos Borromeo.

\section {Conclusions}

After the conquest and colonization by the European population of both territories here considered, in the following decades it slowly began the large-scale 
establishment of small farms and villages, in the case of Lanzarote, and the so-called Indian reductions, in the high mountain plateau of the chilean Andes. 
This was accompanied by the construction of a non negligible number of Christian churches showing the new social and religious situation imposed on those lands .

In the case of Lanzarote, it is possible that in a few places, the orientation of sacred buildings followed the pattern of the aboriginal cult. In others, the
canonical tradition of aligning the temples eastward was respected (with some exceptions to the western quadrant), although with a much greater degree of tolerance than was
usual. Most importantly, we found a statistically significant number of north-northeast oriented churches, which is a notable exception to the rule. After having analyzed
different possibilities to explain this anomaly, we concluded that the most plausible answer lies in the desire to avoid the strong winds prevailing on the island, which come
precisely from that direction, and in particular to minimize the discomfort caused by the sand displaced by the wind on buildings located close or bordering with El Jable.

In the case of the Andean churches, our results show that, unlike what is commonly found in studies involving old European churches \citep{cesar-2014}, just to mention one
example, in the temples we studied we found no single orientation pattern valid throughout the whole region. However, we found that almost half of the churches surveyed have
an orientation that falls within the solar range, with a dominant share in those presenting their altar towards the west. We have also identified some notable cases where the
orientation of the temples seems to obey more to the location of distinctive elements of the terrestrial landscape --as volcanoes or other Aymara culturally relevant
\textit{Apus}-- than to the rising or setting Sun during meaningful dates for the particular dedication of the churches.

\begin{acknowledgement}
I would like to thank my collaborators J.A. Belmonte, C. González-García, A. Guillén, M. Pereira and M.A. Perera-Betancort, for long discussions
on these topics and especially for their continuous help and support. This work was partially financed by CONICET and the University of Buenos Aires.
\end{acknowledgement}


\bibliographystyle{baaa}
\small
\bibliography{aaa-gangui-arqueoastro-baaa2015}
 
\end{document}